\documentclass[12pt]{iopart}

\begin{document}

\title[Persistence of slow dynamics in Tb single ion magnets in conducting polymers ]{Persistence of slow dynamics in Tb(OETAP)$_2$ single molecule magnets embedded in conducting polymers}

\author{T.Orlando,$^{1,2}$ M.Filibian,$^1$ S.Sanna,$^1$ N. Gim\'enez-Agullo,$^3$
C. S\'aenz de Pipa\'on,$^3$ P. Ballester,$^{3,4}$ J.R.
Gal\'an-Mascar\'os,$^3$ P.Carretta $^1$}

\address{$^{1}$ Department of Physics, University of Pavia-CNISM, I-27100 Pavia (Italy)}
\address{$^{2}$ Max Planck Institute for Biophysical Chemistry, D-37077 Gottingen (Germany)}
\address{$^{3}$ Institute of Chemical Research of Catalonia (ICIQ), The Barcelona Institute of Science and Technology, E-43007 Tarragona (Spain)}
\address{$^{4}$ Catalan Institution for Research and Advanced Studies (ICREA), Passeig Lluis Companys 23, E-08010, Barcelona (Spain)}
\ead{pietro.carretta@unipv.it} \vspace{10pt}
\begin{indented}
\item[]May 2016
\end{indented}

\begin{abstract}

The spin dynamics of Tb(OETAP)$_2$ single ion magnets was
investigated by means of muon spin resonance ($\mu$SR) both in the
bulk material as well as when the system is embedded into
PEDOT:PSS polymer conductor. The characteristic spin fluctuation
time is characterized by a high temperature activated trend, with
an energy barrier around 320 K, and by a low temperature tunneling
regime. When the single ion magnet is embedded into the polymer
the energy barrier only slightly decreases and the fluctuation
time remains of the same order of magnitude even at low
temperature. This finding shows that these single molecule magnets
preserve their characteristics which, if combined with those of
the conducting polymer, result in a hybrid material of potential
interest for organic spintronics.

\end{abstract}

\pacs{75.50.Xx, 76.75.+i, 74.50.Gb}
%\maketitle

\section{Introduction}

The trend towards ever-smaller electronic devices is driving
electronics to its ultimate molecular-scale limit leading to
severe drawbacks and to the breakdown of Moore's law \cite{Moore}
while, on the other hand, it allows for the exploitation of
quantum effects both in electronics and in spintronics. Single
Molecule Magnets (SMM) are among the most promising materials to
be used in molecular spintronic devices
\cite{Gatteschi,Bogani,Ardavan} or as logic units in quantum
computers \cite{Leue}, since they combine the classical macroscale
properties of bulk magnetic materials with the advantages of
nanoscale entities, such as quantum coherence. It has already been
shown that molecular magnets can be used to build efficient memory
devices  and, in particular, single crystals can serve as storage
units for dynamic random access memory devices.\cite{Grover} SMMs
can be exploited for all these applications thanks to their
bistability and sizeable energy barriers against magnetization
reversal.\cite{BranzoliJACS} Nevertheless, most of SMMs are
characterized by energy barriers in the tens of K range or below,
giving rise to sufficiently long spin coherence times only at
liquid helium temperature.\cite{Mn12LT} Accordingly, the study of
lanthanide based single ion magnets (SIMs) where the crystal field
(CF) splitting among the $|J, m>$ states gives rise to energy
barriers exceeding 700 K have attracted a great
interest.\cite{Ishikawa1,BranzoliPRB} These lanthanide-based
single-ion magnets (SIMs) were discovered by Ishikawa and
co-workers in a series of double-decker phthalocyanine (Pc)
complexes.\cite{Ishikawa1,Ishikawa2,Ishikawa3} Recently, other
double decker (DD) SIMs \cite{Wang} and
triple-deckers,\cite{Katoh} have been synthesized, with a variety
of ligands including polyoxometallates,\cite{Aldamen}
organometallic compounds,\cite{Tuna}  and organic
radicals.\cite{Lopez} However, all of them show energy barriers
lower than the ones found in Ln(Pc)$_2$ DD, in particular than the
ones observed in Tb(Pc)$_2$. Nowadays, one of the major challanges
is to organize them in monolayers and to address them
individually. Tb(Pc)$_2$ has already been processed on
graphite,\cite{Gondiec} on carbon nanotubes \cite{Klyatskaya} or
on gold.\cite{Katoh2} However, the poor solubility and stability
in the gas phase of these DDs makes it difficult to obtain
perfectly organized arrays from these SIMs. Thus, it would be a
significant breakthrough if analogous magnetic properties could be
found in easy to process single ion magnets.

Tetraazaporphyrins (or porphyrazines) are an attractive
alternative for the bulkier Pc ligands. These macrocycles are
constituted by four pyrrole rings bridged by azanitrogens, giving
an analogous binding mode to Pc but without the extended aromatic
skeleton. This, besides reducing the molecular size, causes an
increase in the solubility and processability of their
corresponding DD complexes. Recently we have reported the
synthesis, structure and the basic magnetic properties of
lanthanide DD complexes with octaethyltetraazaporphyrin
(OETAP).\cite{Nelson} SIM behavior has been found in the Tb and Dy
derivatives, as occurred with the related Pc analogs. Notably, in
this case, the Ln(OETAP)$_2$ neutral complexes are highly soluble
in organic solvents and they can even be sublimed using mild
conditions. Preliminary $^1$H NMR measurements in Tb(OETAP)$_2$
have allowed to evidence an energy barrier of a few hundreds of
Kelvin degrees,\cite{Nelson} smaller than the one of Tb(Pc)$_2$
but still sizeable and interesting for applications. Nevertheless,
the shortening of the NMR relaxation rates lead to the suppression
of the NMR signal over a broad temperature range, not allowing for
an accurate study of the spin coherence time, one of the most
relevant parameters for the technological applicability of  SIMs.
In order to suitably investigate the temperature dependence of the
spin dynamics one can exploit the potential of muon spin resonance
$\mu$SR technique. Here we show that the correlation time for the
spin fluctuations in Tb(OETAP)$_2$ SIMs is characterized by a high
temperature activated behaviour followed by a low temperature
tunneling regime. Moreover, we have investigated the behaviour of
these SIMs when they are embedded in PEDOT:PSS conducting polymers
representing an interesting system where the mutual interplay
between the polymer transport properties and the SIMs magnetic
state can be assessed. We found that the temperature dependence of
the correlation time for the spin fluctuations is very similar to
the one of bulk Tb(OETAP)$_2$, although with a slightly lower
energy barrier, showing that these materials do preserve their SIM
behaviour even when they are embedded into a conducting matrix.

\section{Technical aspects and experimental results}

Tb(OETAP)$_2$ powders were grown and characterized as described in
Ref.\cite{Nelson}. In order to embed the SIMs into the polymer the
following synthesis procedure was adopted. Solid anhydrous 307 mg
of PEDOT:PSS were added to 25 mL of HPLC grade dichloromethane and
the resulting mixture was stirred for 4 days. Then 15.42 mg of
Tb(OETAP)$_2$ were added to the homogeneous suspension an stirred
for 10 min. This solution was dropcast onto a Petri dish open to
air, and the film formed through slow evaporation.

Zero field (ZF) and longitudinal field (LF) $\mu$SR measurements
were carried out at the Paul Sherrer Institut (PSI) - Villigen
(CH) by using Dolly and GPS instruments of the S$\mu$S facility.
In a $\mu$SR experiment a 100\% spin-polarized beam of positive
muons ($\mu^+$) is implanted into the sample. The $1/2$ muon spin
acts as a local magnetic probe and when a magnetic field $B_\mu$
perpendicular to the initial polarization is present, the muon
spin precesses with an angular frequency $\omega_L=\gamma_\mu
B_\mu$, where $\gamma_\mu= 2\pi\times 135.53 $ Mrad/s.T is the
muon gyromagnetic ratio.\cite{Blundell,Yaouanc} When the muon
decays it emits a positron preferentially along its spin
direction, allowing one to reconstruct the time evolution of the
muon spin polarization or, equivalently, of the positron emission
asymmetry $A(t)$. On the other hand, when a longitudinal field
(LF) is applied along the initial muon polarization, no precession
occurs and the time evolution of $A(t)$ is driven by static or
dynamic relaxation mechanisms which are sensitive to the SIMs spin
dynamics.\cite{Blundell,Yaouanc}

%%%%%%%%%%%%%%%%%%%%%%
\begin{figure}[h!]
\vspace{6.8cm} \includegraphics{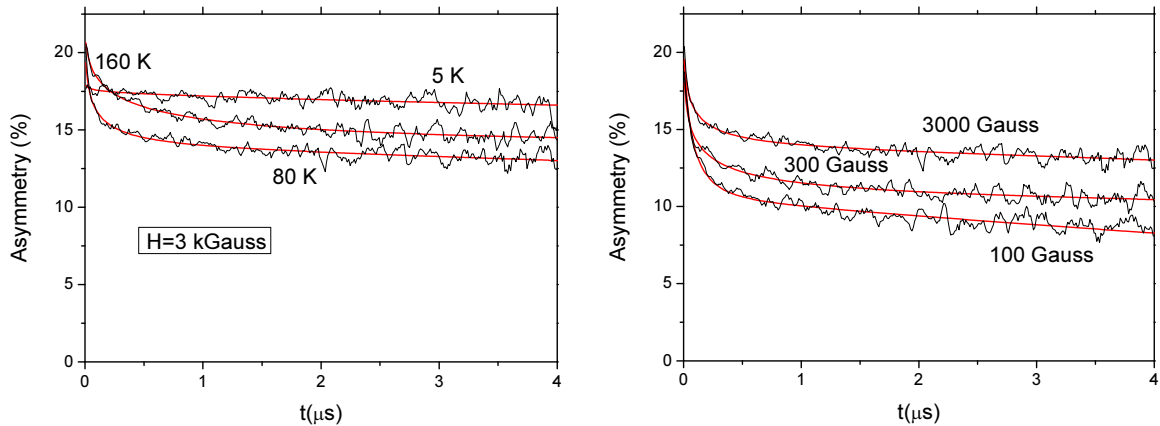} \caption{(left) Time evolution
of the muon asymmetry in Tb(OETAP)$_2$, in a LF of 3 kGauss, at
three different temperatures. The solid lines are the best fits
according to Eq.(1). (right) The asymmetry in the same sample is
now shown, at $T= 80$ K, for three different longitudinal
fields.}\label{asym}
\end{figure}
%%%%%%%%%%%%%%%%%%%%%%%%%%%%%%%%%%%%%%%%%%%%%%%%%%%%%%%%%%%%%%%%%%%%%%%%%%%%%%

Figure~\ref{asym} shows the typical time evolution of the muon
asymmetry in Tb(OETAP)$_2$ in LF geometry. One observes a very
fast initial decay followed by a slow relaxation which indicates a
distribution of relaxation rates, possibly associated with
different muon implantation sites. As the temperature is lowered
the relaxation gets faster and then, below about 80 K, the decay
becomes significantly more stretched and the average rate
decreases again. Accordingly, the asymmetry decay in Tb(OETAP)$_2$
can suitably be fit as the sum of two components
\begin{equation} \label{asymfit}
A(t)= A_1 e^{-(\lambda_F t)^\beta} + A_2 e^{-\lambda t} + Bck\,\,
,
\end{equation}
with $\lambda_F\gg \lambda$ the corresponding relaxation rates,
$\beta$ a stretching exponent and $Bck$ a small constant term
accounting for the sample environment background. Owing to the
very fast initial relaxation the estimate of $\lambda_F$ is
characterized by large errors, even by fixing the amplitude of
$A_1$ and $A_2$ as temperature independent. For this reason we
have rather concentrated on the temperature dependence of the slow
relaxing component relaxation rate $\lambda$.  One notices a clear
maximum around 80 K for the measurements performed at different
magnetic fields (see Fig.3). Moreover, it is observed that upon
increasing the field intensity the asymmetry decay gets
progressively slower. Although this behaviour is considered often
as an evidence of a static field
distribution\cite{Blundell,Yaouanc}, here it is not the case. In
fact, relaxation rates of the order of a few $\mu s^{-1}$ arising
from a static field distribution should be completely quenched in
a magnetic field of 1 kGauss, while they are not (see Fig. 1).
This indicates that the LF relaxation is determined by the
spectral density of the spin fluctuations at $\omega_L$, which is
varied by changing the magnetic field intensity.\cite{Yaouanc}

%%%%%%%%%%%%%%%%%%%%%%
\begin{figure}[h!]
\vspace{6.8cm} \includegraphics{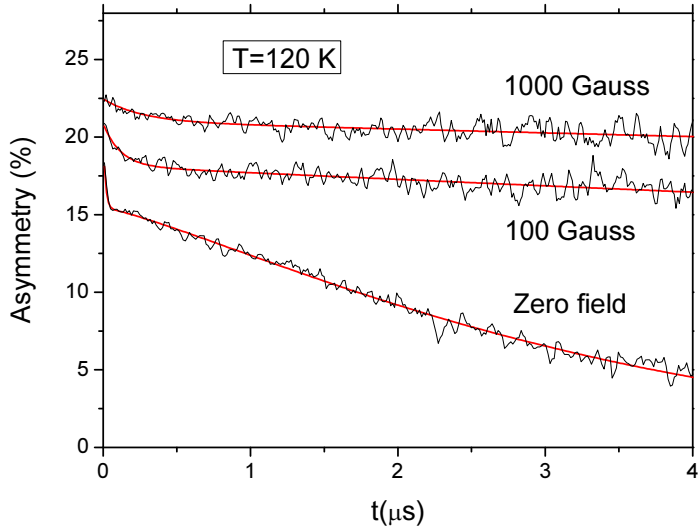} \caption{Time evolution of the muon
asymmetry in Tb(OETAP)$_2$PEDOT:PSS, at $T= 120$ K, for three
different longitudinal fields and in ZF. The solid lines, in LF,
are the best fits according to Eq.(2) in the text, while for the
fit of the ZF data a gaussian decay was added to Eq.(2). }
\end{figure}
%%%%%%%%%%%%%%%%%%%%%%%%%%%%%%%%%%%%%%%%%%%%%%%%%%%%%%%%%%%%%%%%%%%%%%%%%%%%%%

At $H= 3$ kGauss the asymmetry decay could also be fit reasonably
well, even if with a slightly lower accuracy, with a single
stretched exponential decay
\begin{equation} \label{asymstexp}
A(t)= A_s e^{-(\lambda_s t)^\beta} + Bck_s \,\,\, .
\end{equation}
Even if $\lambda_s\gg \lambda$ their temperature dependence is
rather similar, as well as the temperature dependence of the
characteristic spin fluctuation time derived from them (Fig. 4),
as we shall see in the following.

The asymmetry decay in Tb(OETAP)$_2$PEDOT:PSS shows a clearly
different behaviour in zero-field. In fact, as it is shown in
Fig.2 in ZF the asymmetry is characterized by a very fast initial
decay followed by a gaussian-like relaxation. This second
contribution is likely due to the dipolar fields generated by
$^1$H nuclei in the PEDOT:PSS polymer. In fact, this second term
is quenched in a 100 Gauss magnetic field. At those field
magnitude or at larger field intensities the data could be fit by
Eq. 2. The temperature dependence of $\lambda_s$ for
Tb(OETAP)$_2$PEDOT:PSS is shown  in Fig. 3. Similarly to
Tb(OETAP)$_2$, also here one notices a maximum around 80 K.

%%%%%%%%%%%%%%%%%%%%%%
\begin{figure}[h!]
\vspace{6.8cm} \includegraphics{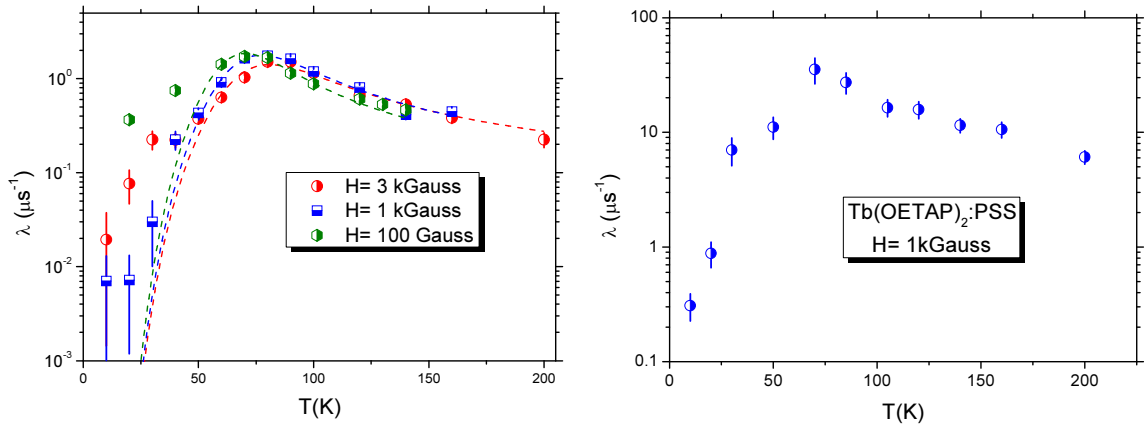} \caption{(left) Temperature
dependence of the LF decay rate $\lambda$ in Tb(OETAP)$_2$ at
three different longitudinal fields. The dashed lines are the best
fits according to Eq.(6) in the text. (right) Temperature
dependence of $\lambda_s$ in Tb(OETAP)$_2$PEDOT:PSS for $H=1$
kGauss. }
\end{figure}
%%%%%%%%%%%%%%%%%%%%%%%%%%%%%%%%%%%%%%%%%%%%%%%%%%%%%%%%%%%%%%%%%%%%%%%%%%%%%%

\section{Discussion}

Ln-based SIMs are characterized by an anisotropy barrier
determined by the crystal field (CF) splitting.  Since the energy
difference between the muon hyperfine levels and the $|J, m>$ CF
levels of Tb$^{3+}$ spin is large, the muon longitudinal
relaxation rate $\lambda$ is driven by an indirect relaxation
mechanism involving a muon spin flip without a change in {\it m}.
This can occur thanks to the tensorial nature of the dipolar
hyperfine coupling constant, which allows the coupling of the
transverse components of the local field ${\it h_{x,y}}$ at the
muon to $\it J_z$. Thus, denoting with ($\tau_m$) the finite
life-time of the crystal field levels induced mainly by
spin-phonon scattering processes, $\lambda$ can be written in the
form\cite{Lasc}
%%%%%%%%%%%%%%%%%%%%%%%%%%%%%%%%%%%%%%%%%%%
\begin{equation}
\vspace{2.5mm} \label{lambda}
      {\lambda}={{\gamma^{2}_{\mu}\langle\Delta h^{2}_\bot\rangle}\over {Z}}\,\sum^{+6}_{m=-6}{\tau_me^{-E_m/T}\,\over
1+\omega^{2}_L \tau^{2}_m}
  \;\;\;\;\;,\vspace{2.5mm}
\end{equation}
$E_m$ being the eigenvalues of the CF levels and $Z$ the
corresponding partition function. It is noted that the low
magnetic field (of the order of 1000 Gauss) applied during $\mu$SR
experiments yield a negligible correction to $E_m$ and, hence, its
effect can be disregarded for $k_BT\gg \mu_BHJ\sim 1$ K. The
life-time for the {\it m} levels can be expressed in terms of the
transition probabilities ${\it p_{m,m \pm 1}}$ between {\it m} and
${\it m\pm 1}$ levels, which depend on the CF eigenvalues and on
the spin-phonon coupling constant C \cite{Villain}:
%%%%%%%%%%%%%%%%%%%%%%%%%%%%%%%%%%%%%%%%%%%%%%%%%%%%%%%%
\begin{equation}
\vspace{2.5mm} \label{taum}
      {1\over \tau_m}=p_{m,m-1}+ p_{m,m+1}
  \;\;\;,
\end{equation}
%\vspace{0.1mm}
\begin{equation}
 \label{taum}
  p_{m , m \pm 1}=
  C
 {
   {(E_{m \pm 1}-E_m)^{3}}
   \over
   {e^{(E_{m \pm 1}-E_m)/T} -1}
 }
 \;\;\;\;
\vspace{2.5mm}
\end{equation}

Since the CF splitting between the lowest levels of Tb$^{3+}$ is
likely larger than the thermal energy, as it will be confirmed in
the following, one can to a first approximation cut the summation
in Eq. 3 to the first term so that one can write
\begin{equation}
\vspace{2.5mm} \label{BPP}
      {\lambda}={{\gamma^{2}_{\mu}\langle\Delta h^{2}_\bot\rangle}}{\tau_c\over
1+\omega^{2}_L \tau_c^{2}}
  \;\;\;\;\;,\vspace{2.5mm}
\end{equation}
with $\tau_c= (1/C\Delta E^3)exp(\Delta E/T)$, where $\Delta E$ is
the CF splitting between the lowest $m=\pm 6$ and $m=\pm 5$
levels. Notice that although a large crystal field splitting
yields a decrease in the fluctuations owing to the Boltzmann
exponential factor, it also implies a large spin-phonon coupling
so that at the end one finds a non-exponential dependence of the
dynamics on the anisotropy barrier. If now one fits $\lambda$ data
of Tb(OETAP)$_2$ in Fig.3 with Eq.\ref{BPP}, taking for $\tau_c$
the activated $T$-dependence reported above, one observes that the
data follow reasonably well the expected trend down to 60-50 K
while clear deviations are found at lower temperature. Hence, at
least down to 60 K $\tau_c$ shows an activated behaviour with an
energy barrier determined by the crystal field splitting. The
failure of the fit at low temperature clearly indicates a
deviation of $\tau_c$ from the activated trend. In order to derive
the temperature dependence of $\tau_c$ from $\lambda(T)$ data, we
inverted Eq.\ref{BPP} obtaining
\begin{equation}
\vspace{2.5mm} \label{tauc}
      {\tau_c}= \frac{1}{\lambda \omega_L}\biggl( \lambda_{max}\pm
      \sqrt{\lambda_{max}^2-\lambda^2}\biggr)
  \;\;\;\;\;,\vspace{2.5mm}
\end{equation}
with $\lambda_{max}$ the maximum value of $\lambda(T)$, achieved
for $\omega_L\tau_c= 1$. Notice that out of the two solutions just
one is physically meaningful, i.e. the one yielding $\tau_c$
increasing on decreasing temperature. The temperature dependence
of $\tau_c$ derived from Eq.\ref{tauc} is shown in Fig.4.

%%%%%%%%%%%%%%%%%%%%%%
\begin{figure}[h!]
\vspace{6.8cm} \includegraphics{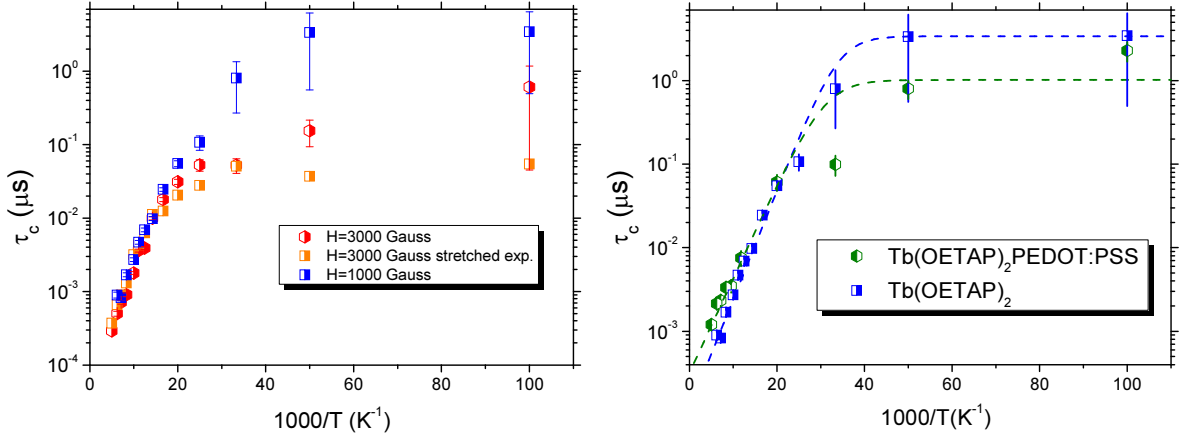} \caption{(left) The correlation time for
the spin fluctuations in Tb(OETAP)$_2$, for different magnetic
field intensities, is reported as a function of $1000/T$. (right)
Comparison of the correlation times of bulk Tb(OETAP)$_2$ with
that of Tb(OETAP)$_2$ embedded in the polymeric matrix as a
function of $1000/T$. The dashed line shows the behaviour
according to Eq. 8, as described in the text.}
\end{figure}
%%%%%%%%%%%%%%%%%%%%%%%%%%%%%%%%%%%%%%%%%%%%%%%%%%%%%%%%%%%%%%%%%%%%%%%%%%%%%%

By plotting $\tau_c$ vs $1000/T$ in a semi-log scale one notices a
high temperature linear trend, namely an activated trend of
$\tau_c$, and then a low temperature flattening which should be
associated with the tunneling among the low-lying energy levels.
In particular, one can write that
\begin{equation}
\vspace{2.5mm} \label{tunn}
      \frac{1}{\tau_c}= \frac{1}{\tau_{tunn}} +  \frac{e^{-\Delta E/T}}{C\Delta E^3}
  \;\;\;\;\;,\vspace{2.5mm}
\end{equation}
with $1/\tau_{tunn}$ the tunneling rate. The high temperature data
for the different magnetic fields can be fit with the same
activation energy $\Delta E= 320 \pm 20$ K  and spin-phonon
coupling constant $C= 250\pm 30$ Hz/K$^3$, a value somewhat
smaller than the one found in LnPc$_2$ compouds
\cite{BranzoliPRB}. On the other hand, the low temperature
behaviour depends on the magnetic field and on the asymmetry
fitting procedure. The dependence on the fitting law stems from
the fact that at low temperature the very fast initial decay gives
rise to a large uncertainty in the relaxation rate when it is fit
with the stretched exponential law.

If now we turn to the behaviour of the correlation time for the
Tb(OETAP)$_2$ spin fluctuations when it is embedded in the
PEDOT:PSS conducting polymers one notices that the correlation
time shows a behaviour very similar to the one found in bulk
Tb(OETAP)$_2$ even if with a slightly lower activation energy
$\Delta E= 260 \pm 30$ K with a spin phonon coupling constant $C=
180\pm 30$ Hz/K$^3$. Remarkably in both materials the fluctuation
time becomes longer than a $\mu$s at temperatures of the order of
a few tens of Kelvin degree, making those materials potentially
interesting for future applications.

\section{Conclusion}

From a series of LF $\mu$SR experiments we have shown that
Tb(OETAP)$_2$ is characterized by a high temperature activated
dynamics with a correlation time increasing above the $\mu$s
already at temperatures of a few tens of K. At low temperature the
tunneling processes among the two fold degenerate ground-state
dominate the spin fluctuations and a clear flattening of the
correlation time is noticed. $\mu$SR measurements performed in
Tb(OETAP)$_2$ molecules show that the spin dynamics is only weakly
affected by the conducting polymers suggesting that these hybrid
materials are potentially interesting for the development of
organic spintronics.

\section*{Acknowledgements}

Hubertus Luetkens and Alex Amato are gratefully acknowledged for
their help during the $\mu$SR measurements at PSI. We acknowledge
financial support from the European Union EU (ERC Stg grant
279313, CHEMCOMP), the Spanish Ministerio de Economia y
Competitividad (MINECO, project CTQ2015-71287-R and the Severo
Ochoa Excellence Accreditation 2014–2018 (SEV-2013-0319), the
Generalitat de Catalunya (2014 SGR 797) and the ICIQ Foundation.

\end{document}